# Estimation of Blood Glucose Level of Type-2 Diabetes Patients using Smartphone Video

Tauseef Tasin Chowdhury, Tahmin Mishma, Md. Saeem Osman, Tanzilur Rahman*

*Department of Electrical and Computer Engineering, North South University, Plot, 15, Block B Kuril - NSU Rd, Dhaka 1229. Phone : 02-55668200, tanzilur.rahman@northsouth.edu**

**Abstract.** This work proposes a smart phone video based approach for the estimation of blood glucose in a non-invasive way. Videos using smartphone camera are collected from the tip of the subject's finger and the frames are subsequently converted into Photoplethysmography (PPG) waveform. Gaussian filter along with Asymmetric Least Square methods have been applied on the PPG signals to remove the high frequency noise, optical and motion interferences. Different signal features such as Systolic and Diastolic Peaks, the time difference between consecutive peaks (DelT), First Derivative peaks, and Second derivative peaks etc have been extracted from the processed signal. Finally, Principal Component Regression (PCR) has been applied for the prediction of glucose level from the extracted features. The proposed model while applied to an unbiased dataset could predict the glucose level with a Standard Error of Prediction (SEP) of 18.31 mg/dL.

## 1. INTRODUCTION

Diabetes mellitus is a chronic incurable disease that causes an array of serious medical complications. The glucose level of diabetes patients need to be regulated by drawing blood several times per day to measure the glucose level, and administer insulin manually [1]. Multiple components like a lancet, one-time test strips and a glucose meter are required to get the estimate. A frequent user such as Type 1 diabetes patient requires to buy and restock these components frequently. The conventional approach of blood glucose monitoring is expensive, uncomfortable and often painful.

Due to inconvenience involved with the conventional approach, the researchers had been in search for a reliable,

accurate and simple method to measure the glucose level without using of selective chemical reagents or physical separations [2, 3]. Over the last two decades, NIR spectroscopy has been established as the most popular method for noninvasive glucose monitoring [4-7]. NIR based technique is a simple technique which has high signal-to-noise ratio as compared to others. Sophisticated data-processing algorithms are required to extract the glucose information in the presence of other dominating signals.

There are other non-invasive techniques which have not been explored enough for this purpose. Photoelectric Plethysmography also known as PPG is a low cost noninvasive technique which measures the volumetric change of blood in the arteries. PPG signal has been widely investigated for the estimation of heart rate by the researchers. Demirezen et al. [8] used Remote Photoplethysmography (rPPG) to detect heart rate from the tissue from the skin whereas Lin et al. [9] did the same form a reflective wrist band used for acquiring PPG. Banerjee et al.[10] proposed a signal processing based approach to estimate coronary artery disease (CAD) from PPG signal. Others [11-14] have demonstrated the potential of PPG signal in analyzing sleep disorders, mental stress, emotion detection. However, PPG waveform which is captured in varying wavelength may contain vital information of other pathological conditions such changes in blood glucose level. Monte-Moreno E [15] demonstrated that PPG signal collected through a pulse-oximiter can be used for the estimation of blood glucose and blood pressure with the help of Machine Learning. The process started with acquiring PPG signal from pulse-oximiter, then they extracted various features from the signal and finally several machine learning techniques were used to estimate the blood glucose and blood pressure from the processed signal. Zhang et al. proposed KNN based binary classification method for identifying diabetic/non diabetic group from smartphone video [16]. The reported approach achieved a very poor accuracy (86.2%), unable to identify pre-diabetic status of the subject which is also vital and unable to remove the dependency of invasive method as the subject still needs to use visit laboratory or use glucometers to know the exact level of glucose which is of utmost importance.

Here, we propose a fully non-invasive technique to predict the glucose level in blood with the help of reflection PPG imaging acquired by smartphone camera. Jonathan et al. [17] previously reported the use of mobile phone camera for reflection PPG imaging from a male subject for the estimation of heart rate. In our study, video data from 18 subjects have been collected from same position of the fingertips and subsequently converted into the respective waveforms. These waveforms have been preprocessed through Gaussian filter for the noise removal. Baseline wandering which is typical in this kind of data acquisition due to motion interferences has been removed using Asymmetric Least Squares (ALS) method. Different types of features have been extracted from the pre-processed PPG signal that includes the Systolic and Diastolic Peaks, the time difference between consecutive peaks (DelT), First Derivative peaks, Second derivative peaks etc. Finally, Principal Component Regression (PCR), a statistical learning approach has been applied for the estimation of glucose level. We believe our model can make blood glucose measurement easier for everyone in the near future.

## 2. PROPOSED METHODOLOGY

An overview of our proposed model has been presented in Figure 1. The process starts with experiment for video data acquisition and conversion into raw PPG signal. After acquiring the PPG signal, different preprocessing techniques were applied to get the useful signal. The preprocessed signal was then used to extract important features relevant to the glucose level and fed into the regression analysis module. The module was trained using the feature set and reference value of glucose taken using the commercially available glucometer (invasive method). Finally, the module was used to predict blood glucose level from unknown sample.

### 2.1. Experiment Details

Consumer grade smartphones available in the market have good camera quality nowadays, which can be a very useful biomedical tool thanks to the advancement of the semiconductor technology. While the quality of acquisition may vary depending on the resolution and number of frames per second, almost all of them can be used for the capture of PPG signal. Videos for such purpose are generally captured from the parts of the body that are closest to the arteries. For example: Earlobe, fingertip etc. In this work, we chose fingertip as it is easier to place on the smartphone camera sensor and the subject does not require any assistance (Figure 2).

A commercially available smartphone (iPhone 7 plus) was used in this study. The imaging unit consists of a WLED as the illumination source next to a 12 megapixel camera at a center-to-center separation of around 5 mm. The phone supports color video recording at about 30 frames per second (fps) at a resolution of 3840 × 2160p and 60 frames per second at 1920 x 1080p. For this study, the fingertip was recorded at 30 fps at a resolution of 720p. This did not reduce signal quality as compared to that of 1080p or 4K but took less phone memory and data transfer time. A color video of the volunteer's index finger placed across both the WLED and camera sensor was recorded in MPEG-4 format onto the phones memory. During the recording the subject was requested to remain at rest to their comfortable position and not to move for as long as possible.

The MPEG-4 files were transferred from the phone memory via USB cable and iTunes software to a laptop computer for further processing. The videos were first converted to RGB JPEG frames and then PPG waveform extraction was performed for red channel signals with custom written Matlab (Mathworks Inc., USA) program. . ROI was not chosen from the frames meaning the average intensity of the whole frame was considered. The PPG signal was extracted as:

$$ppg = MV(f,t) - MV(dc) \ldots \ldots \ldots \ldots (1)$$

Here, MV (dc) is the mean intensity for the entire frame, MV (f, t) is the individual cell mean intensity, f is frame number and t is the time stamp of the frame.

To acquire video data, the tip of the finger was placed on the camera sensor (Figure 2). After that, a video was recorded for about 60 seconds. The duration was selected to give the data some breathing room since there is usually some movement during the placement of the fingertip on the sensor. We simultaneously collected the invasive blood sugar data using a commercial glucose meter (Accu-Check Performa and Accu-Check Active) as a reference. We processed each video by taking the red intensity value first from the whole frame, which gave us multiple values for the whole frame as each section of the frame had a different red intensity value. Then we averaged that value to represent the whole frame. After that we stored each value within an array. The array, when plotted, represented a PPG signal.

Video data was collected from 18 people aged between 15 and 61. Subjects were told to place their fingertip on the camera sensor and wait for 60 seconds. Each video was captured at 30 fps on an iPhone 7 Plus. A total of 5 trials were collected from each. Alongside the video data each subjects age, state of their stomach (Full,

Empty or Partially full) and their invasive glucose level were measured.

## 2.2. Signal Quality

We identified that the best way to collect video data without much motion is to let the subject hold the phone on their hand and then place their fingertip on the camera sensor. It introduces less motion interferences as we naturally hold a phone like this. Other approaches were also tried that caused much movements of the fingers from subject during data collection. A little distortion may result in scattered plots and acquired signals may suffer from baseline variations. An example of such movement is shown Figure 3. Videos were also captured with flash on and off. PPG signals with flash were much cleaner than without Flash which is evident in Figure 4a and 4b. Each frame of the acquired video and the red, green or blue channel intensity values were averaged before storing them into an array for observation purpose. Green and Blue channels are not useful as it gives noisy data (Figure 4c) and didn't resemble a PPG graph. The red channel is useful but needs to be flipped before using (Figure 4d).

Even after careful considerations about the type of camera to be used and orientation of the placement of the fingertip, the acquired signals had certain varieties. Figure 5 provides three sample signals collected from three different subjects, a good type with little to no issue, an average type having little baseline variation and noises and a poor type that has high baseline variation and noises. Therefore a good preprocessing model was required to clean the noises and remove the baseline variations from the acquired and converted signal.

## 2.3. PPG Signal Pre-Processing

For noise removal, various types of digital bandpass filters were used in the literatures. Here, the signals were cleaned with the help of Gaussian smoothing operator that works well as a bandpass filter in the frequency domain. The Gaussian distribution can be explained by the equation below where $c$ is center and $w$ is width:

$$G(x) = e^{-(x-c)^2/2w^2} \ \dots\dots\dots\dots\dots\dots (2)$$

The degree of the smoothing can determined by the adjustment of the standard deviation of the above function. The centroid value c = 0.068 and a width value w = 0.0543 optimized through fine-tuning within a range was identified to best and used for smoothing PPG signals. A sample PPG signal with high frequency noises has been given in Figure 6. The noises could significantly be removed with applying Gaussian Filter. Asymmetric Least Squares method (ALS), a well-known baseline correction technique [18, 19] was chosen for the removal motion interferences. ALS baseline correction method is very useful for correcting data with relatively narrow peaks, which is evident in Figure 7. Here one of sample signal with very high baseline variation has been corrected using ALS. ALS is also quite useful in sharpening the systolic and diastolic peaks which later helps in feature extraction. The baseline correction through ALS improved the signal quality overall as the fluctuations between peaks has been decreased. Some noises however could also be observed signals after processed through ALS. Therefore, we applied ALS correction first, and then Gaussian filter to the ALS corrected signal. A preprocessing model with ALS as a first stage and Gauissian Filter at the second stage significantly improves the quality of the acquired PPG signals of all types. This can be seen in Figure 10 that shows a raw PPG signal processed with Gaussian filters and corrected using ALS.

## 2.4. Feature extraction

Preprocessed PPG signals were subjected for feature extraction. Extracted features were Systolic and Diastolic Peaks, Delt, First Derivative peaks. Each cycle of a PPG signal contains one systolic and one diastolic peak. The locations of these peaks can be extracted as a feature of the PPG signal. Systolic Diastolic peaks found in the preprocessed signal are circled in red in Figure 8. DelT is the time duration between the systolic and diastolic peak in each cycle of the PPG signal, which is marked with red arrows in Figure 8. The processed signal was also converted to its first derivative as it gives higher number of peaks and the peaks are also more pronounced. A 2-point central difference method (eqn 3) was used to determine the first derivative described by Farooq et al [20] and Weng et al [21].

$$d(j) = \frac{(a(j+1) - a(j-1))}{2} \dots \dots \dots (3)$$

The number of peaks doubled from the pre-processed signal after the derivative, which can be seen in Figure 9. After applying the first derivative we used the location of the peaks in each cycle as features.

Since first derivative signals also had negative peaks we had to process the signal further before using. In Figure 10, we can see the raw signal processed through different steps and first derivative peaks that have been detected from the signal.

## 2.5. Regression Analysis

Regression analysis was done on raw signals, preprocessed signals, features extracted from the preprocessed signals to evaluate and compare the performance on the prediction of glucose. In the regression analysis, a pre-defined regression method generates predictions from the input. There are various regression methods like classical least square regression (CLS), principle component analysis (PCA), partial least square regression (PLS), and principal component regression (PCR).

Both PLS and PCR are widely used chemomatric multivariate calibration methods and can be applied when that data set has correlated predictor variables [22]. Both regression methods create new predictor variables (components) as linear combinations of the original predictor variables. PLS creates these components while considering the observed response values. On the other hand, PCR creates component without considering the response values at all. Both regression methods have strong predictive power. PCR has been chosen for this work due to its reported ability to predict glucose level from NIR spectra with a good accuracy [23].

## 3. RESULTS & DISCUSSION

Different PCR models were developed with raw PPG signal without any processing, PPG signal denoised and corrected through ALS and Gaussian filter and $1^{st}$ derivative features extracted from the preprocessed PPG signal. Out of 88 trials, 75% of the trials were used for training and 25% for testing of the models. Figure 11a shows the estimated SEP against different number of principal components (PCs) achieved for PCR model built with different preprocessing approaches. Figure 11b shows the SEP against PCs for PCR model built with RAW signal, preprocessed signal with the approach having low SEP from Figure 11. Here we can see, SEP increased with the increase of No. of principal components for PCR model with raw PPG. SEP on $2^{nd}$ component was 26.1571 mg/dL and on $11^{th}$ component the SEP was 30.4939 mg/dL for this model. A clear reduction of SEP could be observed for all PCs for the other two models. After preprocessing the lowest SEP was recorded to be around 20 mg/dL. The SEP reduced even more (~18 mg/dL) for the PCR model built with 2nd derivative features extracted from the preprocessed PPG. The preprocessing and feature extraction approach have significantly improved the prediction ability of the PCR model.

After applying PCR on waveform processed through different approaches we identified that better accuracy was achieved with 7 PCs. A comparative result received from different approaches while considering 7 PCs is given in Figure 12. It can be seen that SEP for the raw PPG signal is 29.17 mg/dL which has reduced to 19.96 mg/dL due to applying ALS and Gaussian filter on the raw PPG signal. First derivation characteristic points extracted from the preprocessed signal produced the lowest SEP - 18.30 mg/dL when it was used in PCR analysis. Other features like delT did not work well and could achieve an SEP as low as 21.54 mg/dL. After regression analysis, we applied Clarke error grid on the predicted glucose concentration estimated through the model that achieved lowest SEP (preprocessing followed by First derivation characteristics points extraction and then PCR). Clarke error grid is a global method that is used to differentiate between the glucose measurement technique under test and reference measurements [24]. From the Clarke error grid (Figure 13), It could be observed that 82.6% points are in Zone A, 17.4% points in Zone B and no points in the C, D, E. The predicted glucose level also fulfills the current requirement [25] for the clinically acceptable blood glucose monitoring which explains that the measurements should fall within ±15 mg dL−1 for the glucose concentration less than 75 mg dL−1 and ±20% for the concentration greater than 75 mg dL−1. So, we conclude that the PCR model built with preprocessed PPG signal followed by first derivative characteristics points have the ability to estimate glucose in clinically acceptable level.

## 4. CONCLUSION

Here, we have proposed a model comprised of signal processing and regression analysis to estimate the blood glucose level from PPG signal. A smartphone camera based PPG data acquisition and conversion method has been investigated for this purpose. Numerous experiments were done with different acquisition devices and techniques to identify the suitable method of data acquisition. Raw waveforms have been preprocessed and significant feature of signals have been extracted. Finally, PCR algorithm has been applied for the prediction of glucose level and the prediction ability has been validated with the commercial glucometer. The model successfully reduced the SEP to 18.31 mg/dL. First derivative characteristics points played a vital role in reducing the SEP. The experimental results confirm the applicability of the proposed method for glucose level prediction. The proposed model has importance to the community as it can work non-invasively, is a lot easier to use than the traditional methods, can predict glucose value with clinically acceptable accuracy and deliver the glucose level information through smartphone which is now available to most of the people. However, the model has some limitations too. It was trained and tested with small amount of video data. More features need to be investigated to achieve more accurate results. We in future aim at expanding the numbers of subjects, multivariate calibration models, improving the model by finding any irrelevant, scattered data that weakens the prediction model. and implementing the model in the phone for immediate display of glucose level. We believe our model will make blood glucose measurement easier for everyone.

Figure 1: Proposed flow diagram

Figure 2: Basic features of a PPG signal

Figure 3: Intentional finger movement at 20, 30 and 40 seconds

Figure 4: (a) RAW PPG data with flash, (b) Noisy RAW PPG data without flash (c) Green Channel used for plot (d) Flipped Red Channel RAW PPG

Figure 5: Sample RAW PPG signals one from each of the three types. Good, Average and Bad

Figure 6: Raw PPG signal with high frequency noises and the signal after filtered with Gaussian Filter

Figure 7: Systolic and Diastolic Peaks, DelT of the preprocessed PPG

Figure 8: Systolic, Diastolic Peaks, DelT of the preprocessed PPG

Figure 9: First derivative of the preprocessed PPG signal

Figure 10: (a) raw PPG signal (b) preprocessed with ALS (c) preprocessed with Gaussian Filter (d) detection of first derivative peaks from the preprocessed signal

Figure 11: (a) SEP vs. No. of Principal Components of PCR models built after applying different preprocessing steps (b) comparison of PCR models built with raw PPG, preprocessed PPG, features extracted from preprocessed PPG

Figure 12: SEP results achieved through different preprocessing techniques applied on Raw PPG

Figure 13: Clarke error grid on the predicted glucose concentration

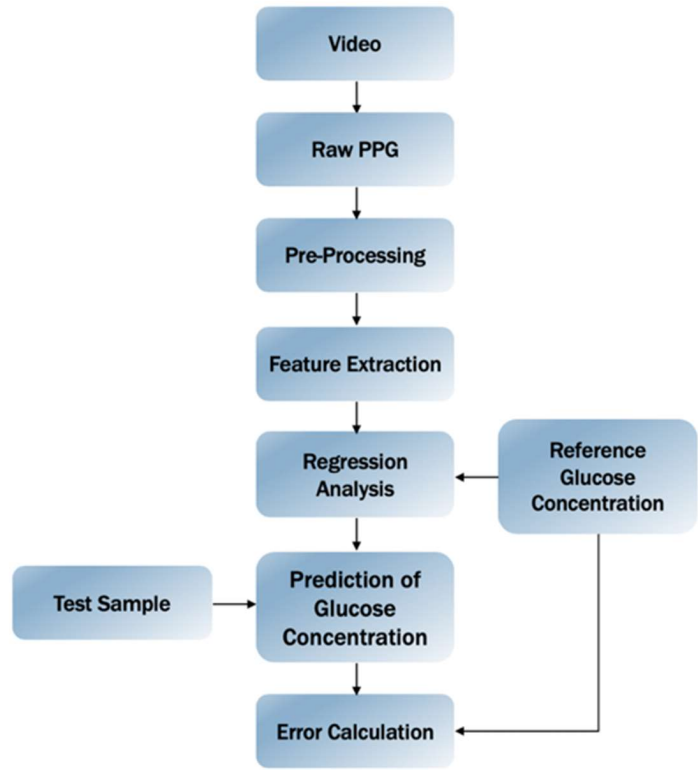

Figure 1

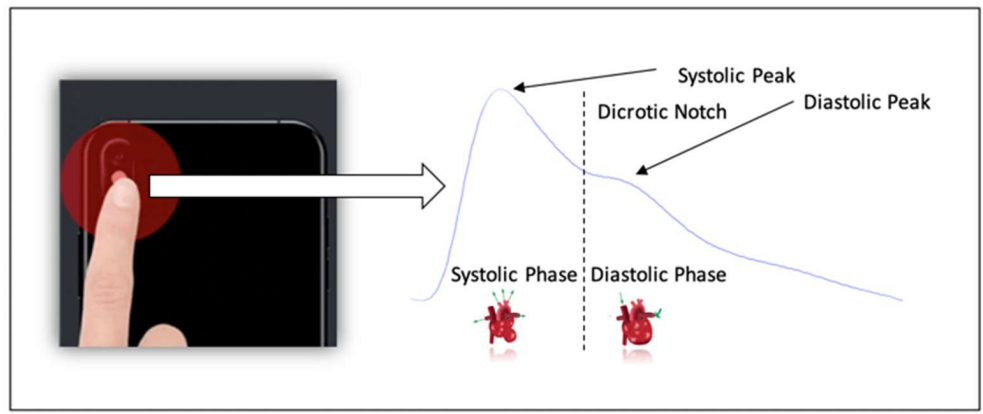

Figure 2

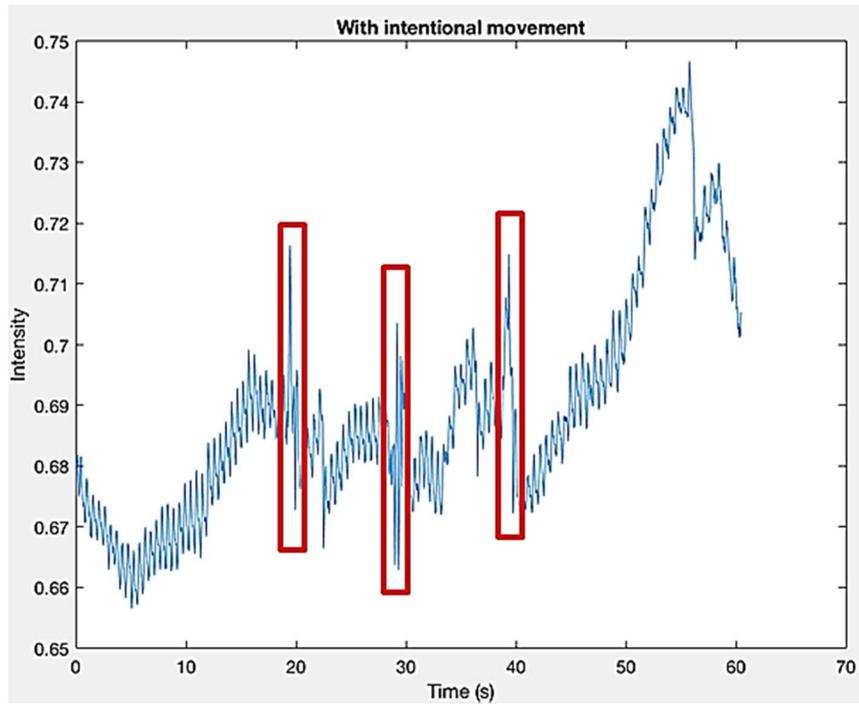

Figure 3

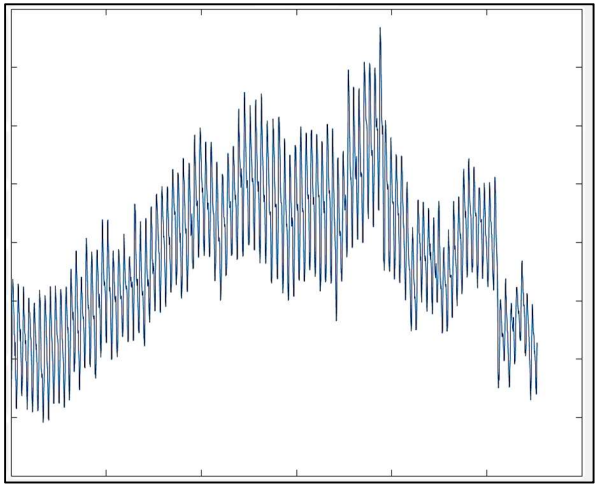

(a)

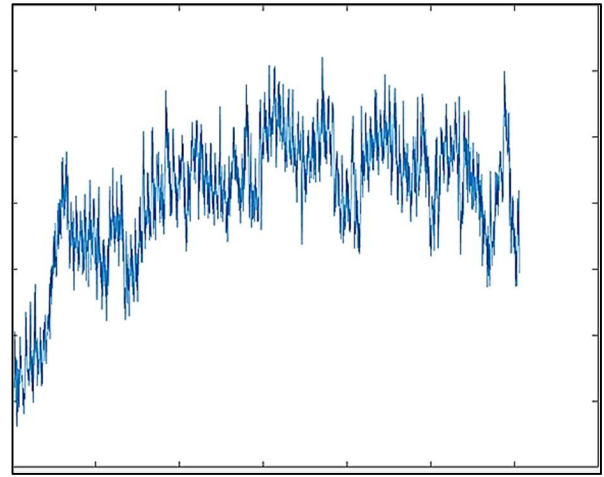

(b)

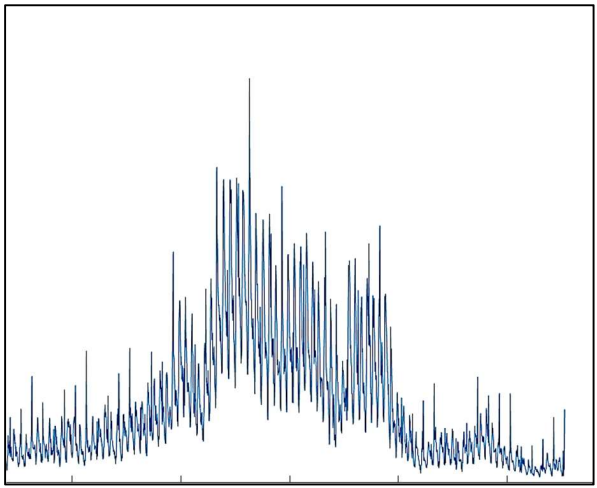

(c)

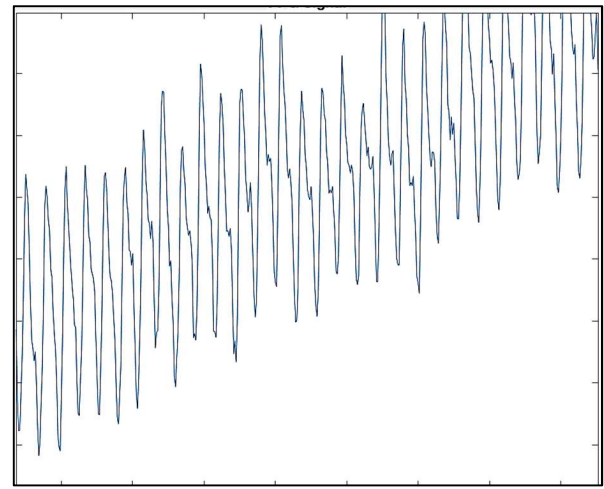

(d)

Figure 4

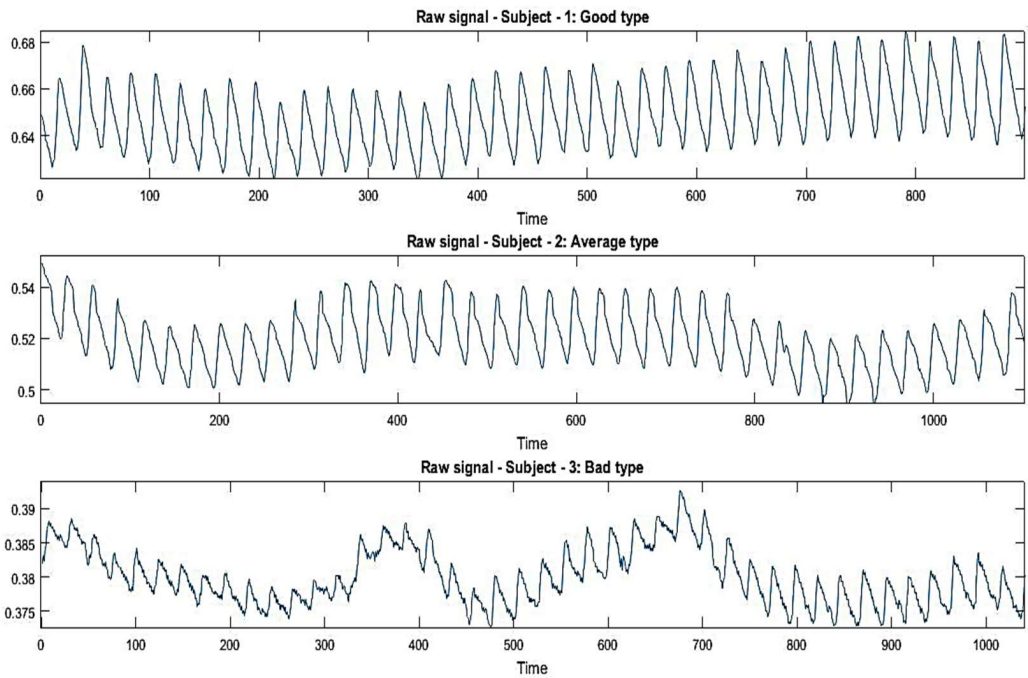

Figure 5

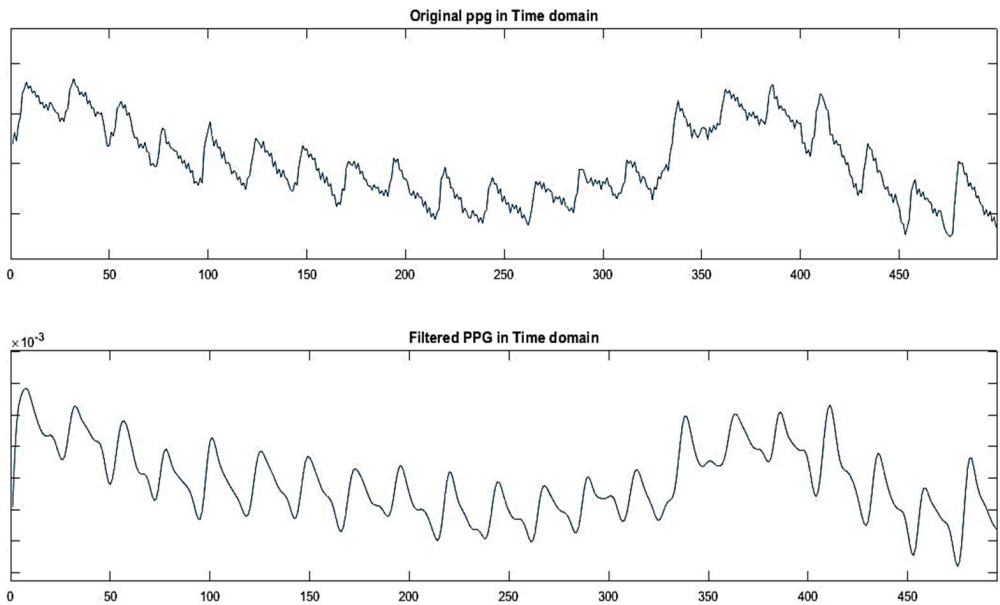

Figure 6

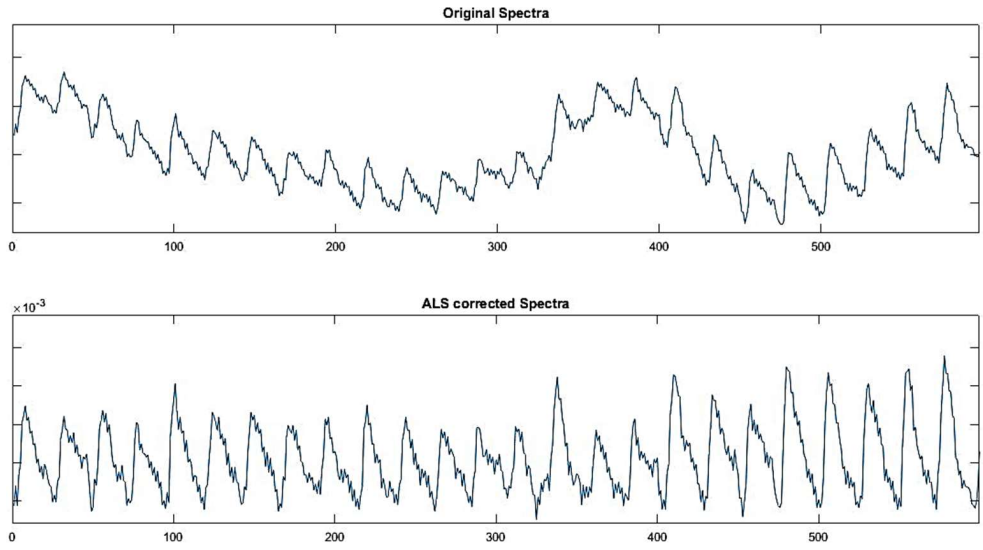

Figure 7

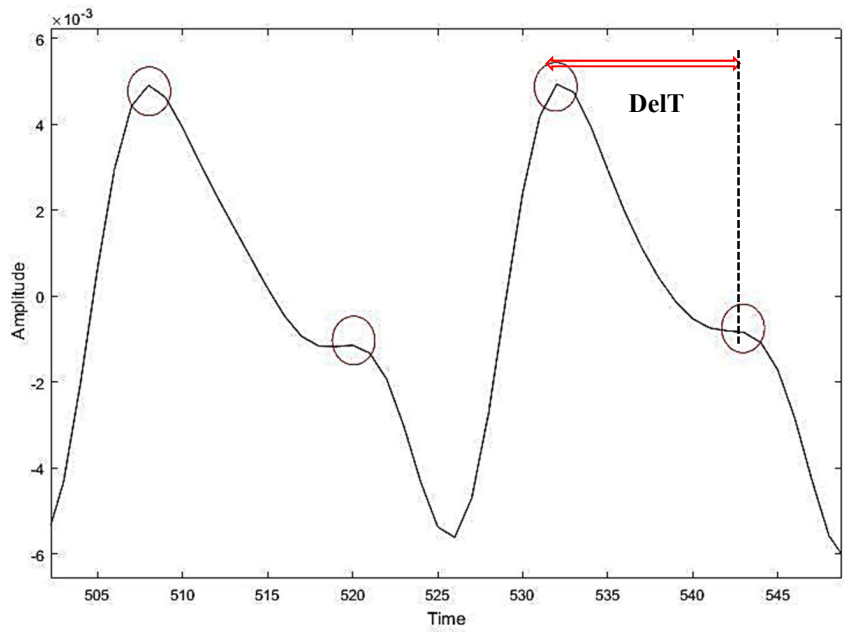

Figure 8

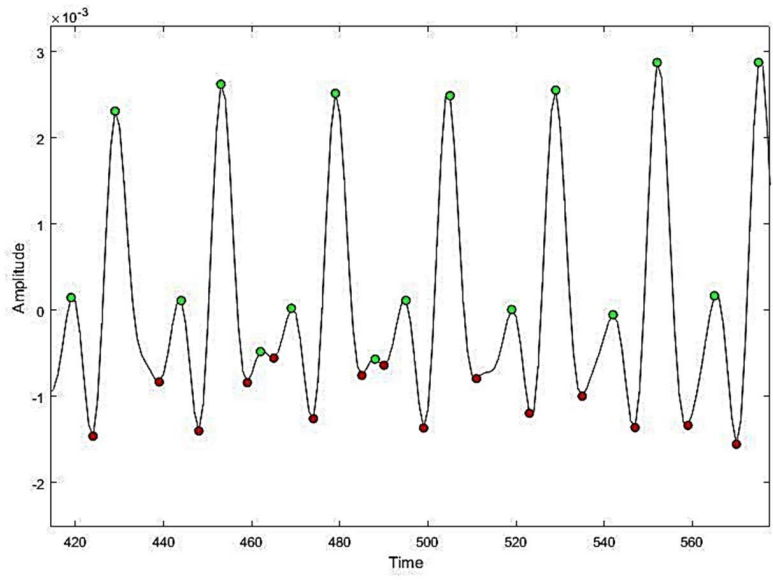

Figure 9

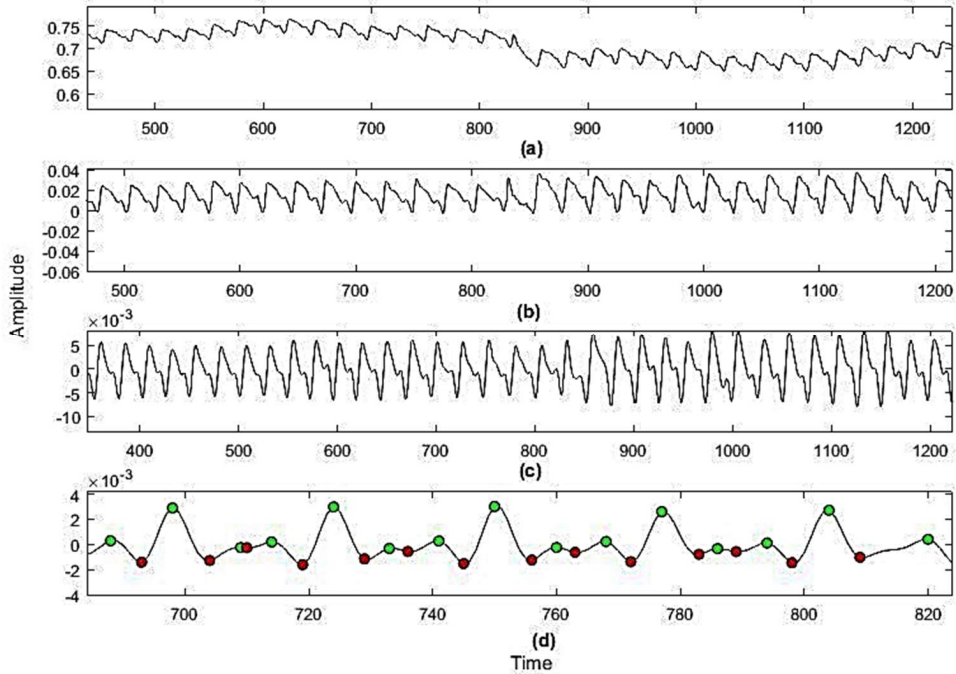

Figure 10

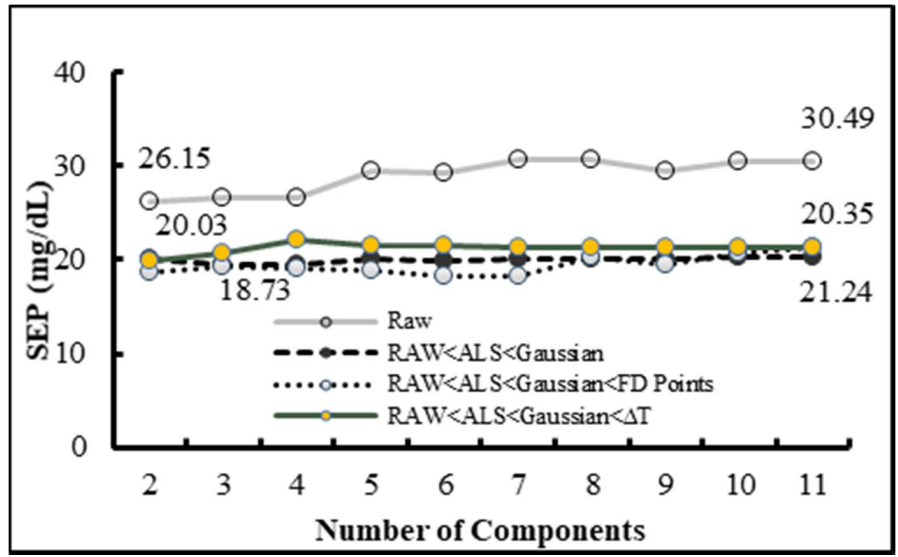

(a)

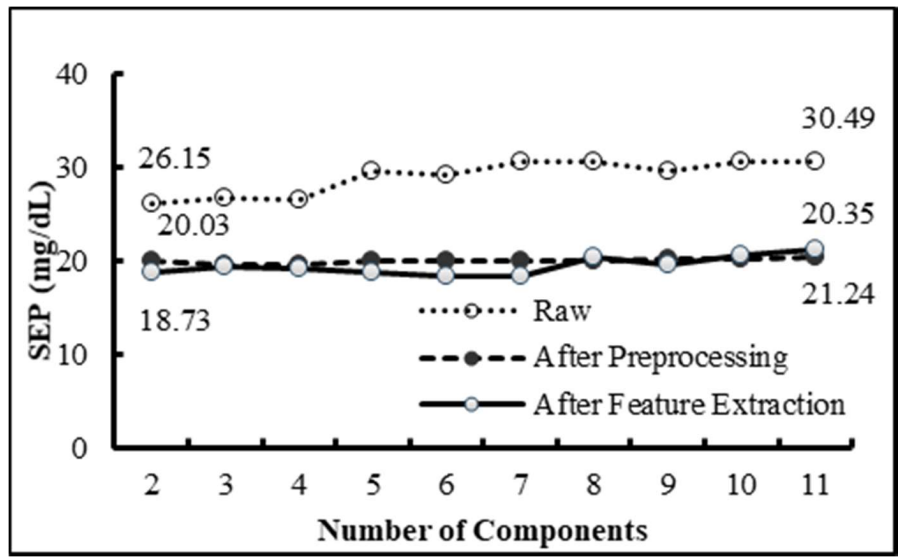

(b)

Figure 11

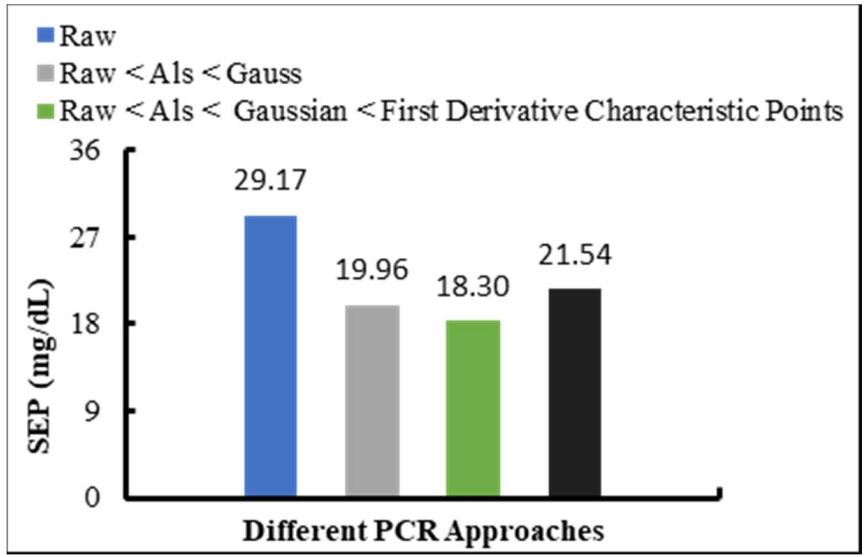

Figure 12

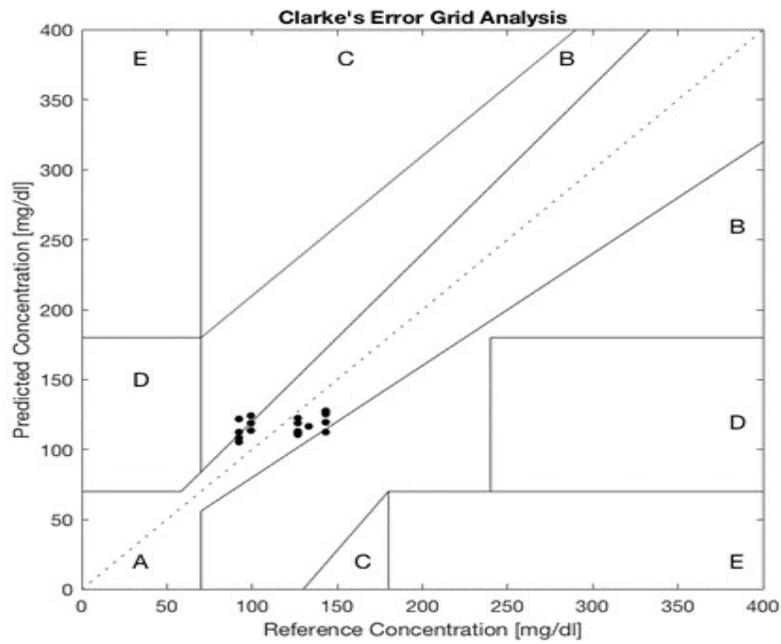

Figure 13